\documentclass[]{fairmeta}

\usepackage{amsmath}
\usepackage{amssymb}
\usepackage{pgfplots}
\pgfplotsset{compat=1.18}

\newif\ifshowcomments
\showcommentstrue

\ifshowcomments
  \newcommand{\jay}[1]{{\small\textcolor{orange}{\bfseries jay: #1}}}
  \newcommand{\todo}[1]{{\textcolor{red}{\bfseries TODO: #1}}}
\else
  \newcommand{\jay}[1]{}
  \newcommand{\todo}[1]{}
\fi

\title{The Case Against Generation for Retrieval: Discriminative Language Models as Effective Retrievers}

\author[1]{Zhe Xu}
\author[1,*]{Prachi Agrawal}
\author[1,*]{Kavosh Asadi}
\author[1,*]{Tianyi Chen}
\author[1,*]{Carl Hu}
\author[1,*]{Justin Johnson}
\author[1,*]{Wuwei Lan}
\author[1,*]{Mingfu Liang}
\author[1,*]{Xi Liu}
\author[1,*]{TIK ON LUI}
\author[1,*]{Oladipo Ositelu}
\author[1,*]{Sandeep Pandey}
\author[1,*]{Ankit Peshin}
\author[1,*]{Feng Qi}
\author[1,*]{Anil Ramakrishna}
\author[1,*]{Kaushik Rangadurai}
\author[1,*]{Frank Shyu}
\author[1,*]{Luke Simon}
\author[1,*]{Yang Yang}
\author[1,*]{Chiyu Zhang}

\affiliation[1]{Meta}
\contribution[*]{Listed in alphabetical order by surname.}

\abstract{Large Language Models (LLMs) have emerged as powerful assets for recommender systems. However, deploying them as generative recommenders or zero-shot rankers at web-scale remains bottlenecked by prohibitive computational overhead and grounding challenges. In this paper, we revitalize the classic, highly efficient two-tower retrieval architecture by adapting LLMs as semantic representation backbones rather than generative engines. We introduce an LLM-native two-tower framework engineered for high-throughput, large-scale retrieval. Our architecture introduces several key innovations: a shared LLM encoder for joint user-item modeling, End-Of-Sentence (EOS) token pooling for compact sequence embedding, cross-dataset transfer learning, knowledge distillation from powerful cross-encoder teachers, and latent reasoning within the user tower. Extensive evaluation across three public benchmarks demonstrates that cross-encoder architecture outperforms current state-of-the-art (SoTA) models, while the efficient two-tower student achieves SoTA-comparable retrieval performance. Furthermore, experiments on internal large-scale production systems yield substantial topline retrieval improvements along with high resilience to model staleness and superior data scaling. Our findings demonstrate that when augmented with modern representation learning, the traditional two-tower paradigm remains an exceptionally competitive and practical solution for industrial retrieval systems.}
\date{\today}

\begin{document}

\maketitle


\section{Introduction}
\label{sec:intro}

Large language models (LLMs) have recently attracted increasing attention in recommender systems due to their strong semantic understanding, instruction-following ability, and broad open-world knowledge \citep{wu2024surveyllmrec,lin2025how}. Existing studies have explored LLMs for multiple recommendation scenarios, including sequential recommendation, conversational recommendation, zero-shot ranking, explainable recommendation, and generative retrieval. A prominent direction is to reformulate recommendation as a language modeling or generation problem, where the model predicts item titles, textual responses, ranked candidate lists, or discrete item identifiers \citep{geng2022recommendation,rajput2023recommender,hou2024large,bao2023tallrec}. For example, P5 unifies several recommendation tasks under a text-to-text language processing framework \citep{geng2022recommendation}, TIGER formulates recommendation as generative retrieval over semantic item IDs \citep{rajput2023recommender}, LLMRank studies LLMs as zero-shot rankers over candidate items \citep{hou2024large}, and TALLRec aligns LLMs with recommendation data through efficient tuning \citep{bao2023tallrec}.

Despite their theoretical promise, fully generative or interaction-heavy LLM-based recommenders encounter severe operational bottlenecks in large-scale retrieval. These architectures incur prohibitive serving costs, rely on latency-inducing autoregressive decoding, and frequently suffer from grounding issues when generated tokens fail to map to valid item identifiers. Crucially, because generative retrieval models emit discrete text tokens rather than explicit item identifiers, the token-to-item translation must occur outside the model. This structural separation causes alignment errors and prediction losses to cascade downstream after inference. In contrast, two-tower (dual-encoder) architectures remain the cornerstone of industrial retrieval. By separately mapping user and item inputs into a shared embedding space, they allow item representations to be precomputed offline and served efficiently via approximate nearest-neighbor (ANN) search \citep{huang2013learning,covington2016deep}. This dichotomy motivates a compelling research question: Can we revitalize and empower classic two-tower architectures with the semantic depth of the LLM era, while preserving their production-proven efficiency and scalability?


In this paper, we study LLM-native two-tower models for recommendation. Moving beyond standard generative paradigms, we adapt pretrained LLMs through supervised fine-tuning on domain-specific data, leveraging them as rich semantic representation backbones for efficient candidate retrieval. Our framework builds on the dual-encoder tradition in semantic matching and retrieval \citep{huang2013learning,reimers2019sentencebert}, while incorporating recent advances in LLM-based text embedding and pooling strategies \citep{wang2024improving,lee2025nvembed}. Specifically, we improve both the cross-encoder teacher and the two-tower student. For the teacher, we use verbalized yes/no relevance scoring and a user-conditioned next-token prediction objective on item text to produce stronger ranking supervision. For the student, we use a shared LLM tower and EOS pooling to align user and item representations in a common semantic space, transfer learning across recommendation datasets, candidate-set score-distribution distillation from the teacher, and Coconut-style latent reasoning in the user tower while preserving precomputable item embeddings \citep{nogueira2019passage,menon2022defense,hou2022towards,hou2023learning,hao2024training}.

We conduct experiments on three real-world recommendation datasets and provide detailed ablation studies. The results show that the proposed cross-encoder enhancements produce a stronger teacher, while the combination of shared LLM encoding, EOS-token pooling, cross-dataset transfer learning, candidate-set score-distribution distillation, and user-tower latent reasoning substantially improves the two-tower student. Our ablation studies further demonstrate the individual contribution of each component. These findings suggest that, even in the era of generative LLM recommenders, two-tower models remain a competitive and practical choice when enhanced with modern LLM-based representation learning and effective training strategies.

The main contributions of this paper are summarized as follows. First, we revisit the classic two-tower recommendation architecture from the perspective of LLM-based representation learning, providing an efficient alternative to fully generative recommendation. Second, we introduce a teacher--student LLM-native retrieval framework that strengthens cross-encoder supervision through verbalized relevance scoring and user-conditioned language modeling, then transfers the teacher's candidate-set ranking preferences to an efficient two-tower retriever. Third, we conduct comprehensive experiments on three real-world datasets with detailed ablation studies to analyze the effectiveness of each proposed component. Fourth, we validate the framework under real-world production conditions, demonstrating exceptional data efficiency (matching classic two tower performance with just 0.5\% of training data), remarkable resilience to model staleness, and superior scaling - offering actionable insights for integrating LLM-native paradigms in industrial retrieval systems.

The remainder of this paper is organized as follows. Section~\ref{sec:related-work} reviews related work on LLM-based recommendation, two-tower retrieval models, transferable recommendation, and knowledge distillation for ranking. Section~\ref{sec:method} introduces the proposed LLM-native two-tower framework and its training strategies. Section~\ref{sec:system} shares the system architecture for serving LLM-Native Two Tower models. Section~\ref{sec:exp} presents the experimental setup, main results, and ablation studies. Section ~\ref{sec:exp internal} shares results on internal production settings and Section~\ref{sec:conclusion} concludes the paper and discusses future directions.

\section{Related Work}
\label{sec:related-work}

Large language models (LLMs) and neural retrieval models provide complementary
foundations for modern recommendation. We give a fuller discussion in
Appendix~\ref{sec:appendix-related-work}.

\subsection{LLM-based Recommender Systems}

LLMs have been applied to recommendation through prompting, instruction
tuning, textual user/item modeling, ranking, and generation
\citep{wu2024surveyllmrec,lin2025how}. Representative generative methods cast
recommendation as language modeling: P5 unifies recommendation tasks in a
text-to-text framework, M6-Rec studies open-ended generative recommendation,
and TIGER generates semantic item identifiers
\citep{geng2022recommendation,cui2022m6rec,rajput2023recommender}. Recent
work also directly generates recommended or ranked items from textual inputs
\citep{ji2024genrec}. At industrial scale, PLUM and the OneRec line extend
generative recommendation through semantic IDs, unified retrieval and ranking,
feedback alignment, and explicit reasoning
\citep{he2025plum,deng2025onerec,zhou2025onerecv2,liu-etal-2026-onerec}.
Other work uses LLMs as zero-shot rankers or instruction-tuned recommenders,
or leverages LLM-generated textual augmentations
\citep{hou2024large,bao2023tallrec,dai2023uncovering,lyu2024llmrec}.

Unlike generative or interaction-heavy LLM recommenders, our work retains
factorized two-tower retrieval for efficient large-scale candidate generation.
We use a shared LLM as a semantic encoder for separate user-history and item
forward passes, then improve the resulting retriever through EOS pooling,
cross-dataset transfer, cross-encoder distillation, and latent reasoning only
in the user tower.

\subsection{Neural Retrieval Models}

Neural retrieval represents queries and candidates in a shared embedding space
and scores them with a lightweight similarity function
\citep{huang2013learning,reimers2019sentencebert,karpukhin2020dense}. In
recommendation, this two-tower factorization allows item embeddings to be
precomputed and indexed, making it effective for candidate generation. It
builds on collaborative-filtering and representation-learning methods such as
matrix factorization, BPR, and Neural Collaborative Filtering
\citep{koren2009matrix,rendle2009bpr,he2017neural}; DSSM, Sentence-BERT, DPR,
and the YouTube recommendation model further established shared encoders for
retrieval and recommendation
\citep{huang2013learning,reimers2019sentencebert,karpukhin2020dense,covington2016deep}.

The efficiency of dual encoders trades off fine-grained user--item
interactions: cross-encoders are stronger but too costly for first-stage
retrieval, while late interaction offers an intermediate alternative
\citep{nogueira2019passage,khattab2020colbert}. Prior work also improves dual
encoders through negative sampling and distillation
\citep{yi2019sampling,wang2021crossbatch,yang2020mixed,menon2022defense}.
Our approach follows this paradigm with a modern shared LLM backbone and
precomputable item representations, while transferring cross-encoder ranking
signals and restricting latent reasoning to the user tower to preserve
retrieval-time factorization.

\section{Preliminaries}
\label{sec:prelim}

We consider a recommendation setting with a user set $\mathcal{U}$ and an item set $\mathcal{I}$. For each user $u \in \mathcal{U}$, let $x_{u}$ denote the textual user input, such as historical interactions, user profile, or preference description. For each item $i \in \mathcal{I}$, let $x_{i}$ denote the textual item input, such as title, description, category, or metadata. The goal is to learn a scoring function $s(u,i)$ such that relevant items receive higher scores and can be retrieved as
\begin{align}
    \mathcal{R}_{u}^{K}
    =
    \operatorname{TopK}_{i \in \mathcal{I}}\, s(u,i),
\end{align}
where $\mathcal{R}_{u}^{K}$ denotes the top-$K$ item set for user $u$.

\subsection{Two-Tower (TT) Models}
\label{sec:prelim-tt}

Two-tower (TT) models are widely used for large-scale retrieval in recommender systems because user and item representations can be computed independently \citep{huang2013learning,covington2016deep,yi2019sampling,yang2020mixed}. Given user and item encoders $f_{\theta_{\mathrm{u}}}$ and $f_{\theta_{\mathrm{i}}}$, the embeddings are computed as
\begin{align}
    \mathbf{z}_{u} = f_{\theta_{\mathrm{u}}}(x_{u})\in \mathbb{R}^{d},
    \qquad
    \mathbf{z}_{i} = f_{\theta_{\mathrm{i}}}(x_{i})\in \mathbb{R}^{d},
\end{align}
The matching score is usually defined by inner product:
\begin{align}
    s_{\mathrm{TT}}(u,i)
    =
    \mathbf{z}_{u}^{\top} \mathbf{z}_{i}.
\end{align}
Since item embeddings can be precomputed, two-tower models support efficient approximate nearest-neighbor retrieval.

%

\subsection{Cross-Encoder (CE) Models}
\label{sec:prelim-ce}

Cross-encoders (CEs) jointly encode the user and item inputs and are commonly used as strong rankers~\citep{nogueira2019passage}. Given a user-item pair, the textual input is constructed as
\begin{align}
    x_{u,i}
    =
    [x_{u}; x_{i}],
\end{align}
and encoded by a textual encoder $f_{\theta}$:
\begin{align}
    \mathbf{h}_{u,i}
    =
    f_{\theta}(x_{u,i}).
\end{align}
The matching score is then computed by a prediction head where rich design space exists; here a single-layer linear transformation is used for illustration:
\begin{align}
    s_{\mathrm{CE}}(u,i)
    =
    \mathbf{w}^{\top}\mathbf{h}_{u,i} + b.
\end{align}
Cross-encoders are more expressive than two-tower models because user and item tokens interact through full self-attention before scoring. However, the representation depends on the specific pair $(u,i)$, so item embeddings cannot be precomputed independently. Therefore, cross-encoders are usually used for ranking or as teacher models for distilling stronger matching signals into efficient retrieval models \citep{menon2022defense}.

\subsection{Contrastive Objective}

Given a user $u$ and an item $i$, let $s(u,i)$ denote the relevance score produced by a matching model. The score function can be instantiated by different backbone models, such as a two-tower model or a cross-encoder.

For implicit-feedback recommendation, models are commonly trained with a contrastive objective. Given a mini-batch of $B$ positive user-item pairs
\begin{align}
    \mathcal{B}
    =
    \{(u_{b}, i_{b}^{+})\}_{b=1}^{B},
\end{align}
we treat the paired item $i_{b}^{+}$ as the positive item for user $u_{b}$, and use the other items in the same mini-batch as in-batch negatives. The candidate set for user $u_{b}$ is therefore
\begin{align}
    \mathcal{C}_{u_{b}}
    =
    \{i_{1}^{+}, i_{2}^{+}, \ldots, i_{B}^{+}\}.
\end{align}
The model defines a softmax distribution over the candidate items:
\begin{align}
    p(i_{j}^{+} \mid u_{b}, \mathcal{C}_{u_{b}})
    =
    \frac{
        \exp(s(u_{b}, i_{j}^{+})/\tau)
    }{
        \sum_{k=1}^{B}
        \exp(s(u_{b}, i_{k}^{+})/\tau)
    },
\end{align}
where $\tau$ is a temperature hyperparameter. The contrastive loss is the cross-entropy loss that encourages each user $u_{b}$ to assign the highest score to its positive item $i_{b}^{+}$:
\begin{align}
    \mathcal{L}_{\mathrm{con}}
    =
    -\sum_{b=1}^{B}
    \log
    \frac{
        \exp(s(u_{b}, i_{b}^{+})/\tau)
    }{
        \sum_{k=1}^{B}
        \exp(s(u_{b}, i_{k}^{+})/\tau)
    }.\label{eq:cl-loss}
\end{align}
This objective is model-agnostic: the same loss applies as long as the backbone produces a scalar relevance score $s(u,i)$ for each user-item pair. To distinguish losses computed from different score functions, we write the score function as an argument when needed; for example, $\mathcal{L}_{\mathrm{con}}(s_{\mathrm{TT}})$ denotes the contrastive loss computed from two-tower scores.
\section{Method}
\label{sec:method}

Our key idea is to transfer the strong ranking capability of cross-encoders into efficient two-tower models through knowledge distillation \citep{hinton2015distilling}. Cross-encoders jointly encode each user-item pair and therefore capture fine-grained interactions, but their pairwise inference cost makes them impractical for large-scale retrieval \citep{nogueira2019passage}. In contrast, two-tower models independently encode users and items, enabling offline item indexing and efficient nearest-neighbor retrieval \citep{covington2016deep,yi2019sampling}. Therefore, we use the cross-encoder as a high-capacity teacher and train the two-tower model as an efficient student.

\begin{figure*}[t!]
    \centering
    \includegraphics[width=.9\textwidth]{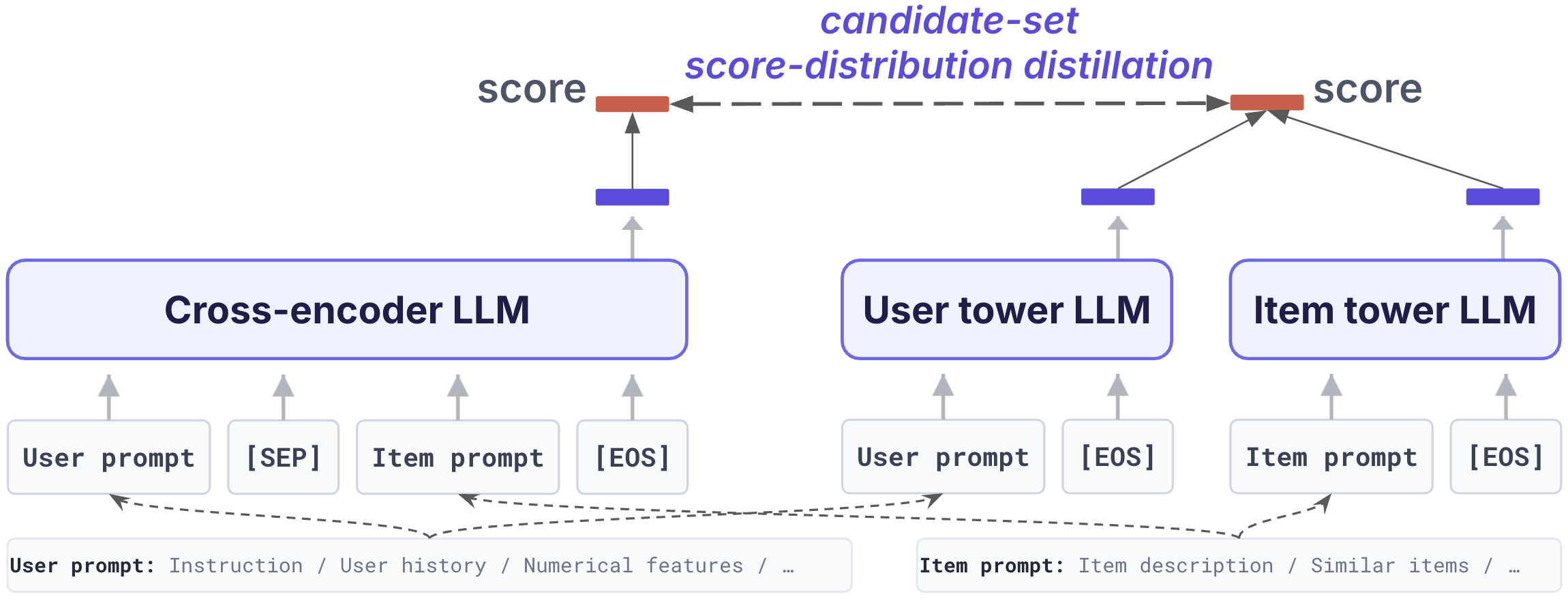}
    \caption{Overview of the proposed LLM-native recommendation framework. The cross-encoder teacher jointly encodes the user and item prompts for a high-quality relevance score, while the two-tower student encodes them separately for efficient retrieval. The dashed arrow denotes distillation, which aligns the teacher and student score distributions over the same candidate set.}
    \label{fig:overview}
\end{figure*}

Our method follows two directions: first, we improve the cross-encoder teacher, which provides the empirical performance upper bound for distillation; second, we improve the distillation effectiveness and generalization ability of the two-tower student, so that it can better approximate the teacher's ranking behavior while preserving retrieval efficiency. An overview of our method is illustrated in Figure~\ref{fig:overview}.

\subsection{Improving the Cross-Encoder Teacher}

The cross-encoder teacher provides the supervision signal for the two-tower student. Therefore, improving the teacher directly improves the quality of the soft labels used in knowledge distillation. In this work, we enhance the cross-encoder teacher from three aspects: using a yes/no output head for relevance prediction, adding an auxiliary next-token prediction loss on item texts, and converting decoder-style backbones into bidirectional encoders.

The techniques are introduced separately but they can be stacked together with minimal modification to show cumulative gains.


\subsubsection{Yes/No Output Head}

Given a user-item pair $(u,i)$, we construct a textual prompt that contains both the user-side text $x_{u}$ and the item-side text $x_{i}$: $
    x_{u,i}^{\mathrm{yn}}
    =
    \operatorname{Prompt}(x_{u}, x_{i}), $
where the prompt asks the model to judge whether item $i$ is relevant to user $u$ and to answer with either ``yes'' or ``no''. Following recent LLM-based ranking designs such as MixLM \citep{li2025mixlm}, we compute the relevance score from the next-token logits of the two verbalized labels.

Specifically, the cross-encoder teacher $f_{\theta}^{\mathrm{CE}}$ takes the full prompt $x_{u,i}^{\mathrm{yn}}$ as input and produces next-token logits:
\begin{align}
    \boldsymbol{\ell}_{u,i}
    =
    f_{\theta}^{\mathrm{CE}}(x_{u,i}^{\mathrm{yn}}).
\end{align}
Let $\ell_{\mathrm{yes}}(u,i)$ and $\ell_{\mathrm{no}}(u,i)$ denote the entries of $\boldsymbol{\ell}_{u,i}$ corresponding to the tokens ``yes'' and ``no'', respectively. We define the cross-encoder relevance score as their logit difference:
\begin{align}
    s_{\mathrm{CE}}(u,i)
    =
    \ell_{\mathrm{yes}}(u,i)
    -
    \ell_{\mathrm{no}}(u,i).
\end{align}
\subsubsection{Auxiliary Next-Token Prediction (NTP) on Item Prompt}

Recommendation data often contains rich item-side textual information, such as titles, descriptions, categories, and reviews. To better adapt the cross-encoder teacher to the recommendation domain, we add an auxiliary next-token prediction objective on item texts. Importantly, this objective is not applied as unconditional language modeling over item texts. Instead, the item text is always predicted conditioned on the corresponding user-side textual input, so that the model learns item semantics in the context of user preference.

For a training pair $(u,i)\in\mathcal{S}$, where $\mathcal{S}$ denotes the training user-item pairs, let the item text be tokenized as $
    x_{i}
    =
    [t_{i,1}, t_{i,2}, \ldots, t_{i,L_i}]. $

Given the user-side text $x_{u}$, the auxiliary next-token prediction loss is defined as
\begin{align}
    \mathcal{L}_{\mathrm{ntp}}
    =
    -\sum_{(u,i)\in\mathcal{S}}
    \sum_{\ell=1}^{L_i-1}
    \log
    p_{\theta}
    \left(
        t_{i,\ell+1}
        \mid
        x_{u}, t_{i,1}, \ldots, t_{i,\ell}
    \right).
\end{align}
This objective follows the standard autoregressive language modeling paradigm used in large language models \citep{brown2020language}, but adapts it to the recommendation setting by conditioning item-text generation on the user context. As a result, the teacher model is encouraged to capture not only item-specific terminology and attributes, but also how these item attributes relate to user preferences.

The final teacher-training objective combines the contrastive objective (Eq.~\eqref{eq:cl-loss}) and the auxiliary user-conditioned next-token prediction loss:
\begin{align}
    \mathcal{L}_{\mathrm{CE}}
    =
    \mathcal{L}_{\mathrm{con}}(s_{\mathrm{CE}})
    +
    \lambda_{\mathrm{ntp}}\mathcal{L}_{\mathrm{ntp}},
\end{align}
where $\lambda_{\mathrm{ntp}}$ controls the strength of the auxiliary objective.

\subsection{Enhancing the Two-Tower Student}

After improving the cross-encoder teacher, we further enhance the two-tower student so that it can better absorb the teacher's ranking knowledge while preserving efficient retrieval. We introduce several techniques: shared user-item encoding, EOS pooling, cross-dataset transfer learning, CE2TT distillation, and Coconut-style latent user-tower reasoning.

Note that we introduce these techniques separately, but they can be stacked together for cumulative gains, as verified empirically.

\subsubsection{Shared User-Item Encoder}

Conventional two-tower models often use separate user and item encoders. In contrast, we use a shared textual encoder $f_{\theta}$ for both user-side and item-side inputs, writing $f_{\theta}(x)$ for its pooled sequence representation:
\begin{align}
    \mathbf{z}_{u} = f_{\theta}(x_{u}),
    \qquad
    \mathbf{z}_{i} = f_{\theta}(x_{i}),
\end{align}
where $x_{u}$ and $x_{i}$ denote the user-side and item-side textual inputs, respectively. Parameter sharing encourages users and items to be embedded into the same semantic space, following the general dual-encoder design used in semantic matching and retrieval \citep{reimers2019sentencebert,huang2013learning}. It also reduces the number of trainable parameters compared with using two independent encoders.

\subsubsection{EOS Pooling}

Given a textual input $x$, the LLM encoder produces a sequence of hidden states: $
    \mathbf{H}_{x}
    =
    [\mathbf{h}_{x,1}, \mathbf{h}_{x,2}, \ldots, \mathbf{h}_{x,L_x}]. $

For decoder-style LLMs, the EOS token is concatenated at the end of the sequence whose embedding is used as the sequence-level representation: $
    \mathbf{z}_{x}
    =
    \mathbf{h}_{x,L_x}. $

EOS pooling is commonly used when adapting decoder-style LLMs into embedding models, since the final token can aggregate information from the preceding sequence under causal attention \citep{wang2024improving,lee2025nvembed}. We empirically found EOS pooling is more effective compared to the mean pooling, detailed in Section~\ref{sec:exp improve tt}.

\subsubsection{Cross-Dataset Transfer Learning}

To improve the generalization ability of the two-tower student, we first mid-train the shared LLM encoder on all available recommendation datasets and then fine-tune it on each target dataset. Suppose we have $R$ datasets, $
    \mathcal{D}_{\mathrm{all}}
    =
    \{\mathcal{D}_{1}, \mathcal{D}_{2}, \ldots, \mathcal{D}_{R}\}. $
The mid-training objective is to minimize
\begin{align}
    \mathcal{L}_{\mathrm{mid}}
    =
    \frac{1}{R}\sum_{r=1}^{R}
    \mathcal{L}_{\mathrm{con}}^{(r)}(s_{\mathrm{TT}}).
\end{align}
After mid-training, the model is further fine-tuned on a specific target dataset by minimizing $\mathcal{L}_{\mathrm{con}}^{(t)}(s_{\mathrm{TT}})$, where $t$ denotes the target dataset. This strategy follows the intuition of transferable recommendation representation learning, where shared behavioral and textual patterns across datasets can improve downstream recommendation performance \citep{hou2022towards,hou2023learning}.

\subsubsection{Cross-Encoder-to-Two-Tower (CE2TT) Distillation}

The cross-encoder teacher produces stronger user-item relevance scores but is too expensive for large-scale retrieval. We therefore distill its ranking behavior into the two-tower student using candidate-set score-distribution distillation \citep{hinton2015distilling}. For each user $u$ and candidate set $\mathcal{C}_{u}$, the cross-encoder teacher produces scores $
    s_{\mathrm{CE}}(u,i),\ i\in \mathcal{C}_{u}. $

The teacher's candidate-set score distribution is defined as $
    q_{\mathrm{CE}}(i\mid u,\mathcal{C}_{u})
    =
    \frac{
        \exp(s_{\mathrm{CE}}(u,i)/T)
    }{
        \sum_{j\in\mathcal{C}_{u}}
        \exp(s_{\mathrm{CE}}(u,j)/T)
    }, $
where $T$ is the distillation temperature. Similarly, the student's candidate-set score distribution is $
    p_{\theta}^{\mathrm{TT}}(i\mid u,\mathcal{C}_{u})
    =
    \frac{
        \exp(s_{\mathrm{TT}}(u,i)/T)
    }{
        \sum_{j\in\mathcal{C}_{u}}
        \exp(s_{\mathrm{TT}}(u,j)/T)
    }. $

The candidate-set score-distribution distillation loss is the KL divergence between the teacher and student distributions:
\begin{align}
    \mathcal{L}_{\mathrm{KD}}
    =
    T^{2}
    \sum_{u}
    \operatorname{KL}
    \left(
        q_{\mathrm{CE}}(\cdot\mid u,\mathcal{C}_{u})
        \,\|\, 
        p_{\theta}^{\mathrm{TT}}(\cdot\mid u,\mathcal{C}_{u})
    \right).
\end{align}
This objective transfers the relative ranking preference of the cross-encoder into the efficient two-tower student.

%
%

\subsubsection{Coconut-Style Latent Reasoning in the User Tower}

User-side inputs often contain multiple historical interactions and preference signals, so a single EOS-pooled representation may not fully summarize the user's intent. Inspired by Coconut-style latent reasoning, which performs reasoning in continuous hidden space rather than through decoded natural-language tokens \citep{hao2024training}, we add one lightweight latent reasoning step only in the user tower. The item tower remains the standard EOS-pooled encoder defined above, i.e., $\mathbf{z}_{i}=f_{\theta}(x_i)$, so item embeddings can still be precomputed. The two towers continue to share the same parameters $\theta$.

Let $g_{\theta}$ denote the shared LLM tower before pooling. For the user side, we first encode the user text without appending EOS and use the last hidden state as a continuous latent token, where $[-1]$ denotes the last hidden state in the sequence:
\begin{align}
    \mathbf{H}_{u}^{(0)}
    =
    g_{\theta}(x_u), \quad
    \mathbf{c}_{u}
    =
    \mathbf{H}_{u}^{(0)}[-1].
\end{align}
The latent token $\mathbf{c}_{u}$ is then appended at the embedding level, followed by a final EOS token. The final EOS hidden state is used as the user representation:
\begin{align}
    \mathbf{H}_{u}^{(1)}
    =
    g_{\theta}([x_u; \mathbf{c}_{u}; \mathrm{EOS}]), \quad
    \mathbf{z}_{u}
    =
    \mathbf{H}_{u}^{(1)}[-1].
\end{align}
Here $\mathbf{c}_{u}$ is never decoded into a discrete word token; it acts as one continuous thought token that gives the user tower an additional step to refine the preference representation. The retrieval score remains the standard two-tower dot product, $
    s_{\mathrm{TT}}(u,i)
    =
    \mathbf{z}_{u}^{\top}\mathbf{z}_{i}, $
where $\mathbf{z}_{i}$ is the unchanged item embedding. This asymmetric computation improves the expressiveness of the user representation while preserving the retrieval efficiency of the item tower.

\subsection{Efficiency Discussion}
\label{sec:efficiency discussion}

We compare the online serving efficiency of our LLM-based two-tower model with generative retrieval methods such as OneRec-Think \citep{liu-etal-2026-onerec}.

For each user query, our model encodes the user prompt once and retrieves items by a dot-product search over precomputed item embeddings. When latent reasoning is enabled, the user tower performs one additional cached decoding step. The subsequent maximum inner-product search can be efficiently parallelized with an approximate nearest-neighbor index.

In contrast, generative retrieval methods must autoregressively decode a sequence that may include a reasoning trace and one or more semantic item identifiers. Given an output sequence of length $N_{\mathrm{gen}}$, this requires a prompt-prefill pass followed by $N_{\mathrm{gen}}-1$ sequential decoding steps. These steps cannot be fully parallelized across output positions and therefore increase serving latency as the generated sequence grows.

Thus, our approach replaces iterative item generation with a single user-side encoding pass and vector retrieval over an offline item index. This design is particularly advantageous when low-latency retrieval over a large catalog is required.

Additional empirical observations and lessons learned from exploring alternative design choices are provided in Appendix~\ref{sec:appendix-observations}.

\section{System Architecture}
\label{sec:system}

\begin{figure}[t!]
    \centering
    \includegraphics[width=.4\linewidth]{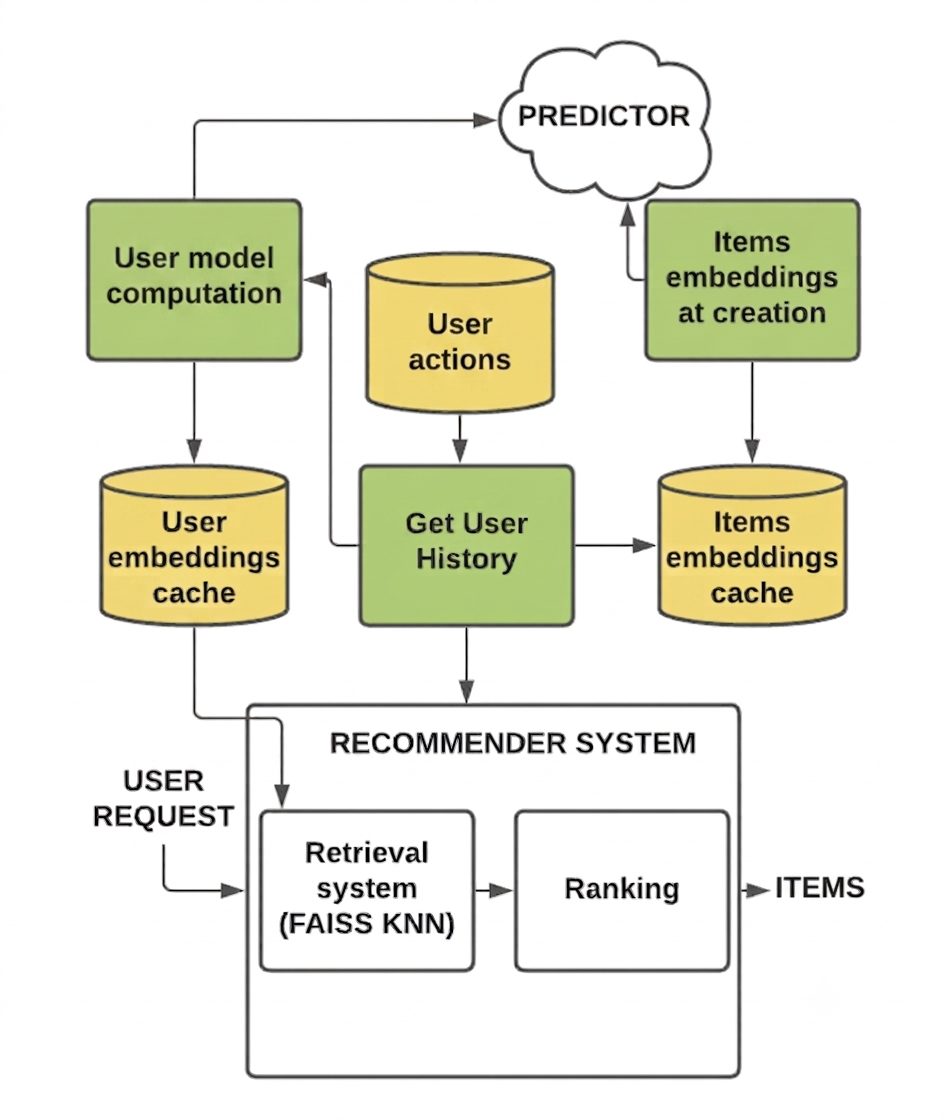}
    \caption{Serving architecture for LLM-Native Two Tower.}
    \label{fig:system arch}
\end{figure}

The system architecture LLM-Native Two Tower for Retrieval is designed to serve highly targeted items at scale while adhering to strict real-time latency requirements. This is achieved through a decoupled, asymmetrical design that isolates intensive language model inference from the online serving path. The architecture is split into three primary components: nearline embedding pipelines for items and users, and a high-performance online retrieval layer.

\subsection{Serving Item Embeddings}
When an item is created or updated, its features are retrieved and integrated into a prompt template for processing. A Predictor service then asynchronously generates a semantic representation of the item. This resulting embedding is indexed in real time, allowing for efficient user-to-item similarity calculations across vast candidate pools during live serving.

\subsection{Serving User Embeddings}
User embeddings are re-computed periodically to reflect fresh engagements within the platform. Similar to the item pipeline, the system aggregates a user's interaction history and processes it through a User LLM. Once updated, the fresh embeddings are cached in a high-throughput, distributed key-value store for rapid online lookup.

\subsection{Online Candidate Retrieval}
The online retrieval layer fetches pre-computed user embeddings from the distributed store within milliseconds. These embeddings are used to query the item index, retrieving relevant candidates in under several hundred milliseconds. To meet these tight latency targets, the engine utilizes embedding space compression to accelerate $k$-nearest neighbor ($k\text{NN}$) computations. The resulting candidate set is then forwarded to downstream ranking stages.

\section{Experiments on Public Datasets}
\label{sec:exp}

\subsection{Setup}

\paragraph{Datasets}

We evaluate on the same three Amazon Reviews benchmarks and
preprocessing protocol used by TIGER \citep{rajput2023recommender}:
\textit{Beauty}, \textit{Sports and Outdoors}, and \textit{Toys and Games}
\citep{he2016ups}. Dataset construction, chronological splitting, and
processed statistics are provided in Appendix~\ref{sec:appendix-datasets}.


\paragraph{Preprocessing.} We convert user and item features into natural language, enabling the LLM tower to more effectively extract key information from recommendation data. The verbalization templates for user and item features are detailed in Appendix~\ref{sec:appendix-preprocessing}.

\paragraph{Backbone Model.}
We use Qwen3-0.6B~\citep{yang2025qwen3} as the backbone for all LLM-based encoders in our framework. We initialize the backbone from the official Qwen3-0.6B checkpoint.\footnote{\url{https://huggingface.co/Qwen/Qwen3-0.6B}}


\paragraph{Metrics}

We evaluate sequential recommendation performance using standard metrics: \emph{Recall@K} and \emph{Normalized Discounted Cumulative Gain (NDCG@K)}. We report $K=5$ and $K=10$, abbreviated as R@K and N@K in the tables. Detailed metric definitions are presented in Appendix~\ref{sec:appendix-metrics}.

\subsection{Results}

\begin{table*}[!t]
\centering
\caption{SOTA comparison on the three Amazon datasets. Baseline results are from Table~1 of ORT (OneRec-Think) \citep{liu-etal-2026-onerec}. OneRec-Think uses a Qwen3-8B backbone, whereas all of our variants use Qwen3-0.6B. The best scores are shown in bold, the second-best scores are underlined, and parenthesized percentages below our scores denote relative changes over ORT.}
\label{tab:sota_comparison}
\newcommand{\ortdiff}[2]{\begin{tabular}[c]{@{}c@{}}#1\\{\small #2}\end{tabular}}
\newcommand{\posdiff}[1]{\textcolor{green!50!black}{(+#1\%)}}
\newcommand{\negdiff}[1]{\textcolor{red!70!black}{(-#1\%)}}
\resizebox{\textwidth}{!}{%
\begin{tabular}{lcccccccccccc}
\toprule
 & \multicolumn{4}{c}{\textbf{Beauty}} & \multicolumn{4}{c}{\textbf{Sports}} & \multicolumn{4}{c}{\textbf{Toys}} \\
\cmidrule(lr){2-5} \cmidrule(lr){6-9} \cmidrule(lr){10-13}
\textbf{Model} & \textbf{R@5} & \textbf{R@10} & \textbf{N@5} & \textbf{N@10} & \textbf{R@5} & \textbf{R@10} & \textbf{N@5} & \textbf{N@10} & \textbf{R@5} & \textbf{R@10} & \textbf{N@5} & \textbf{N@10} \\
\midrule
BERT4Rec      & 0.0232 & 0.0396 & 0.0146 & 0.0199 & 0.0102 & 0.0175 & 0.0065 & 0.0088 & 0.0215 & 0.0332 & 0.0131 & 0.0168 \\
HGN           & 0.0319 & 0.0536 & 0.0196 & 0.0266 & 0.0183 & 0.0313 & 0.0109 & 0.0150 & 0.0326 & 0.0517 & 0.0192 & 0.0254 \\
GRU4Rec       & 0.0395 & 0.0584 & 0.0265 & 0.0326 & 0.0190 & 0.0312 & 0.0122 & 0.0161 & 0.0330 & 0.0490 & 0.0228 & 0.0279 \\
SASRec        & 0.0402 & 0.0607 & 0.0254 & 0.0320 & 0.0199 & 0.0301 & 0.0106 & 0.0141 & 0.0448 & 0.0626 & 0.0300 & 0.0358 \\
TIGER         & 0.0405 & 0.0623 & 0.0267 & 0.0337 & 0.0215 & 0.0347 & 0.0137 & 0.0179 & 0.0337 & 0.0547 & 0.0209 & 0.0276 \\
HSTU          & 0.0424 & 0.0652 & 0.0280 & 0.0353 & 0.0268 & 0.0343 & 0.0173 & 0.0226 & 0.0366 & 0.0566 & 0.0245 & 0.0309 \\
ReaRec        & 0.0450 & 0.0704 & 0.0262 & 0.0344 & 0.0214 & 0.0332 & 0.0116 & 0.0154 & 0.0523 & 0.0764 & 0.0298 & 0.0376 \\
ORT  & \underline{0.0563} & 0.0791 & \textbf{0.0398} & \underline{0.0471} & 0.0288 & 0.0412 & \underline{0.0199} & 0.0239 & 0.0579 & 0.0797 & \underline{0.0412} & 0.0482 \\
\midrule
Ours (TT)     & \ortdiff{0.0473}{\negdiff{16.0}} & \ortdiff{\underline{0.0825}}{\posdiff{4.3}} & \ortdiff{0.0267}{\negdiff{32.9}} & \ortdiff{0.0380}{\negdiff{19.3}} & \ortdiff{\underline{0.0320}}{\posdiff{11.1}} & \ortdiff{\underline{0.0542}}{\posdiff{31.6}} & \ortdiff{0.0193}{\negdiff{3.0}} & \ortdiff{\underline{0.0264}}{\posdiff{10.5}} & \ortdiff{\underline{0.0644}}{\posdiff{11.2}} & \ortdiff{\underline{0.1078}}{\posdiff{35.3}} & \ortdiff{0.0366}{\negdiff{11.2}} & \ortdiff{\underline{0.0506}}{\posdiff{5.0}} \\
\midrule
Ours (CE)     & \ortdiff{\textbf{0.0575}}{\posdiff{2.1}} & \ortdiff{\textbf{0.0957}}{\posdiff{21.0}} & \ortdiff{\underline{0.0351}}{\negdiff{11.8}} & \ortdiff{\textbf{0.0473}}{\posdiff{0.4}} & \ortdiff{\textbf{0.0438}}{\posdiff{52.1}} & \ortdiff{\textbf{0.0677}}{\posdiff{64.3}} & \ortdiff{\textbf{0.0289}}{\posdiff{45.2}} & \ortdiff{\textbf{0.0365}}{\posdiff{52.7}} & \ortdiff{\textbf{0.0829}}{\posdiff{43.2}} & \ortdiff{\textbf{0.1223}}{\posdiff{53.5}} & \ortdiff{\textbf{0.0555}}{\posdiff{34.7}} & \ortdiff{\textbf{0.0682}}{\posdiff{41.5}} \\
\bottomrule
\end{tabular}
}
\end{table*}

\subsubsection{SOTA Comparison}

To compare with state-of-the-art sequential recommendation baselines, Table~\ref{tab:sota_comparison} reports the results from Table~1 of OneRec-Think \citep{liu-etal-2026-onerec} together with our best cross-encoder teacher and the two-tower student. The baselines include BERT4Rec \citep{sun2019bert4rec}, HGN \citep{ma2019hierarchical}, GRU4Rec \citep{hidasi2015session}, SASRec \citep{kang2018self}, TIGER \citep{rajput2023recommender}, HSTU \citep{zhai2024actions}, ReaRec \citep{tang2025think}, and OneRec-Think \citep{liu-etal-2026-onerec}. Notably, OneRec-Think uses Qwen3-8B as its backbone, whereas both of our models use Qwen3-0.6B. Following the source table, ``Sports'' denotes \textit{Sports and Outdoors} and ``Toys'' denotes \textit{Toys and Games}.

\paragraph{Analysis of SOTA Comparison.}
Table~\ref{tab:sota_comparison} shows that our cross-encoder achieves the SOTA overall performance, attaining the best results on 10 of the 12 dataset--metric combinations. Its gains over ORT are particularly pronounced on \textit{Sports} and \textit{Toys}, where R@10 improves by 64.3\% and 53.5\%, respectively. The efficient two-tower student also surpasses ORT in R@10 on all three datasets, with gains of 4.3\%, 31.6\%, and 35.3\% on \textit{Beauty}, \textit{Sports}, and \textit{Toys}, respectively, although its NDCG results are less consistent. These results highlight the complementary roles of the two models: the cross-encoder provides the highest ranking quality, while the two-tower model delivers competitive top-$10$ retrieval performance with an architecture suitable for scalable item retrieval. We next analyze the teacher and student design choices underlying these results.

\subsubsection{Improving Cross-Encoder (CE) Teachers}

\begin{table}[!t]
\centering
\caption{Performance comparison of CE teacher variants across the three Amazon datasets. The best is shown in bold.}
\label{tab:ce_results}
\begin{tabular}{llcccc}
\toprule
\textbf{Dataset} & \textbf{Variant} & \textbf{R@5} & \textbf{R@10} & \textbf{N@5} & \textbf{N@10} \\
\midrule
\multirow{3}{*}{Beauty}
& projection head        & 0.0509 & 0.0862 & 0.0307 & 0.0420 \\
& yes/no head            & 0.0535 & 0.0914 & 0.0328 & 0.0450 \\
& yes/no + NTP & \textbf{0.0575} & \textbf{0.0957} & \textbf{0.0351} & \textbf{0.0473} \\
\midrule
\multirow{3}{*}{Sports}
& projection head        & 0.0391 & 0.0618 & 0.0245 & 0.0317 \\
& yes/no head            & 0.0317 & 0.0550 & 0.0195 & 0.0269 \\
& yes/no + NTP & \textbf{0.0438} & \textbf{0.0677} & \textbf{0.0289} & \textbf{0.0365} \\
\midrule
\multirow{3}{*}{Toys}
& projection head        & 0.0757 & 0.1134 & 0.0492 & 0.0614 \\
& yes/no head            & 0.0814 & 0.1203 & 0.0517 & 0.0643 \\
& yes/no + NTP & \textbf{0.0829} & \textbf{0.1223} & \textbf{0.0555} & \textbf{0.0682} \\
\bottomrule
\end{tabular}
\end{table}

\paragraph{Analysis of Cross-Encoder Variants.}
Table~\ref{tab:ce_results} isolates the contributions of the output head and the auxiliary NTP objective. The yes/no head improves over the projection head on \textit{Beauty} and \textit{Toys and Games}, but degrades all four metrics on \textit{Sports and Outdoors}, indicating that the benefit of verbalized relevance scoring alone is dataset-dependent.

Adding the user-conditioned NTP objective resolves this inconsistency: the yes/no+NTP variant achieves the best score on all 12 dataset--metric combinations. In particular, relative to the yes/no head, it raises R@10 from $0.0914$ to $0.0957$ on \textit{Beauty}, from $0.0550$ to $0.0677$ on \textit{Sports}, and from $0.1203$ to $0.1223$ on \textit{Toys}; the corresponding N@10 improvements are $0.0023$, $0.0096$, and $0.0039$. These consistent gains suggest that predicting item text conditioned on user context supplies complementary recommendation-specific supervision. We therefore use the yes/no+NTP configuration as the cross-encoder teacher for distillation.

\subsubsection{Improving Two-Tower (TT) Students}
\label{sec:exp improve tt}

Table~\ref{tab:tt_ablation_results} reports a leave-one-out ablation of the proposed two-tower student. All variants use a shared encoder with EOS pooling; each ablation removes one of transfer learning (TL), cross-encoder-to-two-tower distillation (CE2TT), or Coconut-style latent reasoning \citep{hao2024training} in the user tower from the full model. Results for every combination of these three optional components, together with the vanilla two-tower and Shared+EOS-only baselines, are reported in Appendix~\ref{sec:appendix-tt-ablation}.

\begin{table}[!t]
\centering
\caption{Leave-one-out ablation of the TT student across the three Amazon datasets. All variants use Shared+EOS; each ablated variant removes one component from the full model. The best score in each dataset--metric pair is shown in bold.}
\label{tab:tt_ablation_results}
\begin{tabular}{llcccc}
\toprule
\textbf{Dataset} & \textbf{Variant} & \textbf{R@5} & \textbf{R@10} & \textbf{N@5} & \textbf{N@10} \\
\midrule
\multirow{4}{*}{Beauty}
& Full & \textbf{0.0473} & \textbf{0.0825} & 0.0267 & \textbf{0.0380} \\
& w/o TL & 0.0431 & 0.0812 & 0.0241 & 0.0363 \\
& w/o CE2TT & 0.0427 & 0.0715 & 0.0252 & 0.0345 \\
& w/o Latent & 0.0463 & 0.0794 & \textbf{0.0269} & 0.0374 \\
\midrule
\multirow{4}{*}{Sports}
& Full & \textbf{0.0320} & \textbf{0.0542} & \textbf{0.0193} & \textbf{0.0264} \\
& w/o TL & 0.0300 & 0.0515 & 0.0174 & 0.0243 \\
& w/o CE2TT & 0.0258 & 0.0417 & 0.0160 & 0.0211 \\
& w/o Latent & 0.0308 & 0.0530 & 0.0185 & 0.0257 \\
\midrule
\multirow{4}{*}{Toys}
& Full & 0.0644 & \textbf{0.1078} & 0.0366 & \textbf{0.0506} \\
& w/o TL & 0.0604 & 0.1050 & 0.0346 & 0.0490 \\
& w/o CE2TT & 0.0626 & 0.0992 & \textbf{0.0383} & 0.0502 \\
& w/o Latent & \textbf{0.0650} & 0.1063 & 0.0370 & 0.0503 \\
\bottomrule
\end{tabular}
\end{table}

\paragraph{Analysis of Two-Tower Ablations.}
CE2TT is the most consequential optional component: removing it reduces R@10 by $0.0110$ ($13.3\%$), $0.0125$ ($23.1\%$), and $0.0086$ ($8.0\%$) on \textit{Beauty}, \textit{Sports}, and \textit{Toys}, respectively, relative to the full model. TL and latent reasoning provide smaller but complementary gains; removing either lowers R@10 on every dataset. Consequently, the full model achieves the best R@10 and N@10 on all three datasets. A few top-$5$ metrics favor an ablated variant, but the full configuration offers the most consistent top-$10$ retrieval quality.

\section{Experiments on Internal Datasets}
\label{sec:exp internal}

We evaluate the LLM-native retrieval framework on large-scale internal data under production serving conditions. We compare it with a heavily tuned \textbf{DLRM Two-Tower} production retriever following standard DLRM practices~\cite{naumov2019deep}. This baseline uses independent user and item towers whose dot product supports efficient vector search, with most parameters concentrated in embedding layers that accommodate item-ID vocabularies of up to $O(10{M})$ entries.

\subsection{Setup}
\label{sec:internal-setup}

\paragraph{Data.}

We evaluate on internal retrieval system. Each example is serialized as a structured sequence of user, item, and task blocks. Continuous features are discretized with Finite Scalar Quantization (FSQ) \citep{mentzer2023finite}; legacy dense embeddings are compressed into a fixed set of learnable query tokens with a Q-Former
\citep{li2023blip2}; and ID/event sequences are verbalized into short textual snippets. This common representation permits new feature blocks to be added without changing the model architecture. Unless otherwise noted, models use an $8$K-token context.

\paragraph{Models.}
Unless otherwise stated, the internal models use the same Qwen3-0.6B backbone
as the public experiments and perform prefill-only discriminative scoring. We evaluate a two-tower model for retrieval and a cross-encoder provides distillation targets for the former. For capacity
scaling, we additionally evaluate a $4$B dense backbone and a Mixtral-style
sparse mixture-of-experts model with $8$ experts and top-$2$ routing. We refer to our internal system as \textbf{LLM-native TT/CE} and to the production baseline as \textbf{DLRM}.

\paragraph{Metrics.}
Our primary quality metric is evaluation \textbf{Normalized Entropy (NE)}, for
which lower is better. We report relative \textbf{NE gain (\%)}, defined as the
percentage reduction in NE against the stated baseline. We measure
generalization by the NE change on evaluation days $ds{+}1, ds{+}2,\dots$ after
the training cut-off, and report serving throughput (QPS) for efficiency.

\subsection{Results and Learning}
\label{sec:internal-results}

\subsubsection{LLM-native Two-Tower Retrieval}

\paragraph{NE parity at reduced data scale.}
\textbf{LLM-native TT matches the NE of the production DLRM baseline while using only 0.5\% of the training data.} Under the same evaluation, LLM-native CE improves NE by $+2.25\%$ over the baseline, indicating headroom for the retriever through CE-to-TT distillation.

\paragraph{Segmentation analysis.}
We further evaluate the LLM-native TT across user and item engagement segments. \textbf{It improves NE by $+5.5\%$ on tail items}---the bottom $15\%$ by engagement---relative to DLRM, consistent with the potential value of semantic representations when interaction histories are sparse. \textbf{LLM-native TT also improves NE by $+2.3\%$ for head users}---the top $12\%$ by engagement---suggesting that its sequence modeling captures additional signal in long interaction histories.

\paragraph{Serving efficiency.}
Table~\ref{tab:internal-efficiency} summarizes the serving optimizations for LLM-native TT. To reduce the long inputs caused by digit-level tokenization of numerical features, we quantize continuous features into $64$ FSQ bins and inject their learnable embeddings as soft tokens. This compression increases QPS by $19.7\%$ and improves NE by $0.3\%$. We also prune $3$ of $28$ transformer layers selected by a search-proxy metric and recover quality through knowledge transfer, which increases QPS by $10.6\%$. Static vocabulary pruning removes rarely used embedding entries to reduce memory footprint and increases QPS by $2.0\%$. Finally, post-training FP8 quantization increases QPS by $13.6\%$ with an approximately $0.003\%$ NE regression.

\begin{table}[t]
\centering
\caption{Improving serving efficiency for LLM-native TT without hurting NE}
\label{tab:internal-efficiency}
\begin{tabular}{ll}
\toprule
Technique & QPS Gain \\
\midrule
Numerical Features FSQ Compression & 19.7\% \\
Depth pruning & 10.6\% \\
Static vocabulary pruning  & 2.0\% \\
Post-training Quantization & 13.6\% \\
\bottomrule
\end{tabular}
\end{table}

\subsubsection{LLM-native Cross-Encoder Generalization and Scaling}
\paragraph{Generalization under staleness.}
Unlike classic DLRMs, which demand continuous retraining to prevent rapid performance decay, LLM-native models demonstrate superior resilience to model staleness. To evaluate this property, we trained both a DLRM baseline and an LLM-native model on data up to date $ds$ and benchmarked their frozen inference performance over three subsequent days ($ds+1$, $ds+2$, and $ds+3$) without parameter updates. All evaluation metrics are reported relative to a daily-recurrently retrained, fresh production DLRM baseline. As shown in Table~\ref{tab:internal-staleness}, the frozen DLRM exhibits severe quality degradation over time (losing $-2.68\%$ on $ds+2$ and $-4.30\%$ on $ds+3$) due to its reliance on shifting, ephemeral item ID distributions. Conversely, the \textbf{Frozen LLM-Native CE} maintains robust performance across future dates (+2.25\% on $ds+1$, +2.21\% on $ds+2$, and +2.23\% on $ds+3$) by grounding representations in a stable token vocabulary.
\begin{table}[t]
\centering
\caption{Staleness. Both DLRM and LLM-Native model is trained until $ds{+}0$ and evaluated on 
$ds{+}1$/$ds{+}2$/$ds{+}3$ (negative =
degradation).}
\label{tab:internal-staleness}
\begin{tabular}{llll}
\toprule
Model & $ds{+}1$ (NE) & $ds{+}2$ & $ds{+}3$ \\
\midrule
DLRM (production) & baseline & baseline & baseline \\
Frozen DLRM & baseline & $-2.68\%$ & $-4.30\%$ \\
Frozen LLM-Native CE     & $+2.25\%$ & $+2.21\%$ & $+2.23\%$ \\
\bottomrule
\end{tabular}
\end{table}

\paragraph{Data scaling.}
While the LLM-Native TT retriever achieves NE parity with the production DLRM baseline using only 0.5\% of the training data, the sample efficiency and scaling trajectory of the LLM-Native CE model are even more compelling. LLM-Native CE reaches baseline NE parity utilizing merely 0.15\% of the training data. Furthermore, as dataset volume scales by $2\times$ and $3\times$, LLM-Native CE demonstrates compounding performance gains (+1.59\% and +2.19\% NE improvement over its $1\times$ baseline), outperforming the scaling curve of its classic DLRM counterpart (+1.30\% and +1.80\%, as detailed in Table~\ref{tab:internal-datascaling}). We attribute this superior scaling behavior to the stable semantic vocabulary of LLMs, which enables continuous, cumulative feature learning, in contrast to the fragile, ephemeral item ID reliance of traditional DLRM models.

\begin{table}[t]
\centering
\caption{Data scaling. The model performance (NE) is the same at $1\times$ data while LLM-Native CE shows NE gains at  $2\times$/$3\times$ relative to that model's own
$1\times$ point.}
\label{tab:internal-datascaling}
\begin{tabular}{lll}
\toprule
Model & $2\times$ & $3\times$ \\
\midrule
DLRM (production)   & $+1.30\%$ & $+1.80\%$ \\
LLM-native CE       & $+1.59\%$ & $+2.19\%$ \\
\bottomrule
\end{tabular}
\end{table}

\paragraph{Model scaling and capacity.}
To explore the upper limits of our architecture's capacity, we conducted a headroom analysis by scaling model capacity and compute along three distinct axes: ($i$) increasing the parameter size of the backbone LLM, ($ii$) integrating Mixture-of-Experts (MoE) layers, and ($iii$) expanding computation via latent reasoning. As detailed in Table~\ref{tab:internal-capacity}, each strategy yields substantial NE improvements. While these scaled configurations exceed our current serving budget and were omitted from the primary benchmark results, they demonstrate significant performance headroom when additional computational resources become available.

\begin{table}[t]
\centering
\caption{Headroom Study: Gains from Model Scaling}
\label{tab:internal-capacity}
\begin{tabular}{ll}
\toprule
Variant & NE gain \\
\midrule
LLM-Native CE $0.6\text{B}$ & baseline \\
+ Latent reasoning (residual-stream)          & $+0.41\%$ \\
+ Mixtral MoE ($8$ experts, top-$2$)                & $+0.91\%$ \\
LLM-Native CE $4\text{B}$ & $+1.90\%$ \\
\bottomrule
\end{tabular}
\end{table}




\section{Conclusion}
\label{sec:conclusion}

This paper studies how LLMs can improve point-wise recommendation architectures, with a focus on cross-encoders and efficient two-tower retrieval models. We develop stronger LLM-based cross-encoders and show that their fine-grained matching capability can be effectively transferred to two-tower models through distillation, preserving efficient retrieval. Across three public benchmarks, our methods consistently improve both model families and demonstrate the value of cross-encoder supervision for retrieval. Large-scale industrial experiments further show that the proposed LLM-native two-tower model is competitive with a heavily tuned production DLRM baseline while using only a small fraction of the training data, and provides gains across challenging traffic segments while remaining robust to model staleness. Overall, our results suggest that combining discriminative LLM representations with factorized architectures offers a practical and cost-effective path toward web-scale recommendation.

\clearpage

\bibliographystyle{assets/plainnat}
\bibliography{ref}

\clearpage

\beginappendix
\section{Data Preprocessing}
\label{sec:appendix-preprocessing}
As discussed in the main text, we preprocess the data to improve LLM performance on recommendation tasks. The user-side input contains two parts:
\begin{itemize}
    \item Recent items (last 5): each item is formatted as ``[Brand] Title (LeafCategory, L2Category) Price Rating''.
    \item Taste summary (for users with at least 3 history items): top-3 L2 categories with percentages (only those at least 15\%), top-2 brands, and price range (min--max).
\end{itemize}
The item-side input contains:
\begin{itemize}
    \item Core information: ASIN, title, leaf category, L2 category, brand, price, and average rating count.
    \item Description snippet: the first 100 characters of the product description, truncated at a word boundary.
    \item Also-bought signal: up to 3 co-purchased items formatted as ``Also bought: [Brand] Title (LeafCat) | [Brand] Title (LeafCat) | ...''.
\end{itemize}

\section{Public Dataset Details}
\label{sec:appendix-datasets}

We use three public benchmarks from the Amazon Product Reviews dataset
\citep{he2016ups}: \textit{Beauty}, \textit{Sports and Outdoors}, and
\textit{Toys and Games}. These datasets contain user reviews and item
metadata collected from May 1996 to July 2014 and are commonly used for
sequential recommendation evaluation.

Following TIGER \citep{rajput2023recommender}, we construct a chronological
item sequence for each user by sorting the user's review history by timestamp
and remove users with fewer than five reviews. We use a leave-one-out protocol:
the last item in each sequence is used for testing, the second-to-last item
for validation, and the remaining items for training. In our pointwise setup,
the chronological item sequence serves as the user-side information.

\begin{table*}[!h]
\centering
\caption{Statistics of the three processed Amazon review datasets.}
\label{tab:dataset_statistics}
\begin{tabular}{lrrrr}
\toprule
Dataset & \# Users & \# Items & Mean Len. & Median Len. \\
\midrule
Beauty & 22,363 & 12,101 & 8.87 & 6 \\
Sports and Outdoors & 35,598 & 18,357 & 8.32 & 6 \\
Toys and Games & 19,412 & 11,924 & 8.63 & 6 \\
\bottomrule
\end{tabular}
\end{table*}

\section{Metrics}
\label{sec:appendix-metrics}

\paragraph{Recall@K.}
Let $\mathcal{U}$ denote the set of users, $i_u^{\ast}$ the ground-truth next item for user $u \in \mathcal{U}$, and $\mathcal{R}_u^{K}$ the set of top-$K$ items ranked by the model.
Recall@K is defined as
\begin{equation}
\mathrm{Recall}@K
= \mathbb{E}_{u\in\mathcal{U}}\left[
\mathbb{I}\!\left( i_u^{\ast} \in \mathcal{R}_u^{K} \right)\right],
\end{equation}
where $\mathbb{I}(\cdot)$ is the indicator function.

\paragraph{NDCG@K.}
For user $u$, the Discounted Cumulative Gain at $K$ (DCG@K) is computed as
\begin{equation}
\mathrm{DCG}@K
= \sum_{j=1}^{K}
\frac{2^{\mathrm{rel}_{u,j}} - 1}{\log_2(j+1)},
\end{equation}
where $\mathrm{rel}_{u,j} \in \{0,1\}$ indicates whether the item ranked at position $j$ matches the ground-truth next item $i_u^{\ast}$.

The Ideal DCG at $K$ (IDCG@K) is obtained by placing the ground-truth item at the first position. Under the standard leave-one-out evaluation protocol for sequential recommendation, this value is always $1$: $\mathrm{IDCG}@K
= \frac{1}{\log_2(1+1)} = 1$. NDCG@K is then defined as
\begin{equation}
\mathrm{NDCG}@K
= \mathbb{E}_{u\in\mathcal{U}}\left[
\frac{\mathrm{DCG}@K}{\mathrm{IDCG}@K}\right].
\end{equation}


\section{Implementation Details}
\label{sec:impl}

\subsection{Backbone and Hardware}
Training uses \texttt{bfloat16} mixed precision, the AdamW optimizer
(weight decay $0.01$), and a linear warmup over the first $10\%$ of steps
followed by cosine decay. Gradient checkpointing is enabled throughout to
reduce activation memory. All experiments use NVIDIA A100-SXM4 80\,GB GPUs;
the cross-encoder and the two-tower pre-training/fine-tuning stages use $4$
GPUs, and the distillation stage uses $8$ GPUs. Text is tokenized with
the Qwen3 tokenizer, truncating queries to $512$ and items to $128$ tokens.

\subsection{Cross-Encoder (Yes/No Head + NTP Loss)}
\label{sec:impl-ce}
The contrastive and NTP losses introduced in the main text are jointly
optimized as $\mathcal{L}_{\mathrm{CE}} = \mathcal{L}_{\mathrm{con}}(s_{\mathrm{CE}}) + \lambda_{\mathrm{ntp}}\mathcal{L}_{\mathrm{ntp}}$,
with $\lambda_{\mathrm{ntp}}=0.5$. Hyperparameters are listed in
Table~\ref{tab:ce-hparams}.

\begin{table}[t]
\centering
\caption{Cross-encoder (yes/no head + NTP loss) hyperparameters.}
\label{tab:ce-hparams}
\begin{tabular}{ll}
\toprule
Hyperparameter & Value \\
\midrule
Negatives per positive & $31$ \\
NTP loss weight $\lambda_{\mathrm{ntp}}$ & $0.5$ \\
Epochs & $10$ \\
GPUs & $4\times$ A100-80GB \\
Per-GPU batch & $32$ \\
Effective batch & $128$ \\
Learning rate & $2\times10^{-5}$ \\
Weight decay & $0.01$ \\
Warmup ratio & $0.1$ \\
Max length (query / item) & $512$ / $128$ \\
Precision & bf16 \\
Seed & $42$ \\
\bottomrule
\end{tabular}
\end{table}

\subsection{Two-Tower Retriever}
\label{sec:impl-tt}
During CE$\to$TT distillation, the contrastive loss and candidate-set
score-distribution distillation loss introduced in the main text are jointly optimized as
$\mathcal{L}_{\mathrm{TT}} = \mathcal{L}_{\mathrm{con}}(s_{\mathrm{TT}}) + \lambda_{\mathrm{KD}}\mathcal{L}_{\mathrm{KD}}$,
with $\lambda_{\mathrm{KD}}=0.2$. Table~\ref{tab:tt-hparams} reports the
remaining implementation-specific settings for the two-tower retriever.

\begin{table}[t]
\centering
\caption{Two-tower retriever hyperparameters (shared encoder, EOS pooling,
transfer $+$ CE$\to$TT distillation $+$ latent reasoning).}
\label{tab:tt-hparams}
\begin{tabular}{ll}
\toprule
Hyperparameter & Value \\
\midrule
Embedding & raw $1024$-d \\
Similarity / temperature $\tau$ & dot product / $0.07$ \\
\midrule
\multicolumn{2}{l}{\emph{Pre-training \& fine-tuning (transfer)}} \\
GPUs & $4\times$ A100-80GB \\
Per-GPU batch & $32$ \\
Epochs & $10$ \\
Learning rate & $2\times10^{-5}$ \\
\midrule
\multicolumn{2}{l}{\emph{CE$\to$TT distillation}} \\
GPUs & $8\times$ A100-80GB \\
Per-GPU batch & $32$ \\
In-batch negatives & $31$ \\
Teacher candidates $K$ & $50$ \\
Epochs & $5$ \\
Learning rate & $1\times10^{-5}$ \\
KD loss weight $\lambda_{\mathrm{KD}}$ & $0.2$ \\
Distill temperature $\mathcal{T}$ & $1.0$ (Beauty) / $0.5$ (Sports, Toys) \\
\midrule
\multicolumn{2}{l}{\emph{Shared}} \\
Optimizer & AdamW (wd $0.01$, warmup $0.1$) \\
Max length (query / item) & $512$ / $128$ \\
Precision & bf16 \\
Seed & $42$ \\
\bottomrule
\end{tabular}
\end{table}

\section{Full Two-Tower Ablation}
\label{sec:appendix-tt-ablation}

The main paper reports a leave-one-out ablation of the three optional
two-tower components. Table~\ref{tab:tt_ablation_full_results} provides
results for every combination of these components. The vanilla baseline uses separate encoders with
mean pooling; every other variant uses a shared encoder with EOS pooling. TL,
CE2TT, and Latent denote transfer learning, cross-encoder-to-two-tower
distillation, and Coconut-style latent reasoning in the user tower,
respectively.

\begin{table*}[!t]
\centering
\caption{Complete two-tower ablation across the three Amazon datasets. The best score in each dataset--metric pair is shown in bold.}
\label{tab:tt_ablation_full_results}
\begin{tabular}{lcccccccc}
\toprule
\textbf{Dataset} & \textbf{Shared+EOS} & \textbf{TL} & \textbf{CE2TT} & \textbf{Latent} & \textbf{R@5} & \textbf{R@10} & \textbf{N@5} & \textbf{N@10} \\
\midrule
\multirow{9}{*}{Beauty}
& - & - & - & - & 0.0214 & 0.0389 & 0.0125 & 0.0180 \\
& \checkmark & - & - & - & 0.0415 & 0.0672 & 0.0253 & 0.0337 \\
& \checkmark & - & \checkmark & - & 0.0429 & 0.0800 & 0.0241 & 0.0360 \\
& \checkmark & \checkmark & - & - & 0.0430 & 0.0714 & 0.0260 & 0.0351 \\
& \checkmark & \checkmark & \checkmark & - & 0.0463 & 0.0794 & \textbf{0.0269} & 0.0374 \\
& \checkmark & - & - & \checkmark & 0.0425 & 0.0711 & 0.0251 & 0.0343 \\
& \checkmark & - & \checkmark & \checkmark & 0.0431 & 0.0812 & 0.0241 & 0.0363 \\
& \checkmark & \checkmark & - & \checkmark & 0.0427 & 0.0715 & 0.0252 & 0.0345 \\
& \checkmark & \checkmark & \checkmark & \checkmark & \textbf{0.0473} & \textbf{0.0825} & 0.0267 & \textbf{0.0380} \\
\midrule
\multirow{9}{*}{Sports}
& - & - & - & - & 0.0054 & 0.0109 & 0.0033 & 0.0050 \\
& \checkmark & - & - & - & 0.0205 & 0.0347 & 0.0131 & 0.0177 \\
& \checkmark & - & \checkmark & - & 0.0298 & 0.0510 & 0.0175 & 0.0243 \\
& \checkmark & \checkmark & - & - & 0.0261 & 0.0426 & 0.0166 & 0.0219 \\
& \checkmark & \checkmark & \checkmark & - & 0.0308 & 0.0530 & 0.0185 & 0.0257 \\
& \checkmark & - & - & \checkmark & 0.0223 & 0.0387 & 0.0132 & 0.0184 \\
& \checkmark & - & \checkmark & \checkmark & 0.0300 & 0.0515 & 0.0174 & 0.0243 \\
& \checkmark & \checkmark & - & \checkmark & 0.0258 & 0.0417 & 0.0160 & 0.0211 \\
& \checkmark & \checkmark & \checkmark & \checkmark & \textbf{0.0320} & \textbf{0.0542} & \textbf{0.0193} & \textbf{0.0264} \\
\midrule
\multirow{9}{*}{Toys}
& - & - & - & - & 0.0294 & 0.0489 & 0.0175 & 0.0237 \\
& \checkmark & - & - & - & 0.0600 & 0.0968 & 0.0367 & 0.0487 \\
& \checkmark & - & \checkmark & - & 0.0614 & 0.1049 & 0.0353 & 0.0493 \\
& \checkmark & \checkmark & - & - & 0.0563 & 0.0928 & 0.0332 & 0.0450 \\
& \checkmark & \checkmark & \checkmark & - & \textbf{0.0650} & 0.1063 & 0.0370 & 0.0503 \\
& \checkmark & - & - & \checkmark & 0.0604 & 0.0946 & \textbf{0.0383} & 0.0493 \\
& \checkmark & - & \checkmark & \checkmark & 0.0604 & 0.1050 & 0.0346 & 0.0490 \\
& \checkmark & \checkmark & - & \checkmark & 0.0626 & 0.0992 & \textbf{0.0383} & 0.0502 \\
& \checkmark & \checkmark & \checkmark & \checkmark & 0.0644 & \textbf{0.1078} & 0.0366 & \textbf{0.0506} \\
\bottomrule
\end{tabular}
\end{table*}

\section{Additional Observations}
\label{sec:appendix-observations}

We also evaluated several plausible variants that were not retained in the final configuration because they did not provide consistent gains. These observations further motivate the design choices used throughout the paper.

\begin{itemize}
    \item \textbf{Sharing the user and item encoder is beneficial.}
    We evaluated separate LLM encoders for the user and item towers, which increase model capacity and allow tower-specific parameters. In our experiments, however, the shared encoder consistently performed better, suggesting that parameter sharing more effectively aligns user and item representations in a common semantic space.

    \item \textbf{Embedding matching is less effective.}
    In addition to CE2TT distillation, we matched the cross-encoder and two-tower EOS representations with an $L_2$ loss on positive user-item pairs. This embedding-level objective degraded retrieval performance relative to candidate-set score-distribution distillation, indicating that matching the teacher's relative preferences over candidates is more useful than directly aligning their pooled representations.

    \item \textbf{Additional latent steps do not improve retrieval quality.}
    Adding latent reasoning to the item tower did not improve performance, likely because item-side inputs are comparatively simple. Likewise, using more than one latent step in the user tower substantially enlarged the computation graph and increased memory use without measurable gains. We therefore use a single latent step only in the user tower.

    \item \textbf{Decoder-to-encoder conversion is not consistently helpful.}
    We also evaluated a Dec2Enc variant that removes the causal attention mask to enable bidirectional token interactions. Although this modification is plausible for discriminative representation learning, it did not yield consistent improvements and often degraded both the cross-encoder and two-tower models. We therefore exclude it from the final method.
\end{itemize}

\section{Extended Related Work}
\label{sec:appendix-related-work}

\subsection{LLM-based Recommender Systems}

Large language models (LLMs) have recently been widely explored for recommender systems due to their semantic understanding, instruction-following ability, and open-world knowledge \citep{wu2024surveyllmrec,lin2025how}. Existing work incorporates LLMs into recommendation through prompt-based recommendation, generative recommendation, zero-shot ranking, textual user/item representation, explanation generation, and conversational recommendation.

A representative direction formulates recommendation as a language modeling problem. P5 unifies multiple recommendation tasks under a text-to-text pretraining and prompting framework \citep{geng2022recommendation}, while M6-Rec studies generative pretrained language models as open-ended recommender systems \citep{cui2022m6rec}. More recently, generative retrieval methods such as TIGER represent items with semantic identifiers and train generative models to decode target item IDs \citep{rajput2023recommender}. Other works further investigate LLMs that directly generate recommended items or ranked outputs from textual inputs \citep{ji2024genrec}.

Recent industrial systems further extend generative recommendation to production-scale retrieval. PLUM adapts pretrained language models through semantic item IDs, continued pretraining, and task-specific fine-tuning for industrial-scale generative recommendation \citep{he2025plum}. At Kuaishou, OneRec unifies retrieval and ranking in an end-to-end generative framework; subsequent work improves the approach with a lazy decoder-only architecture and alignment with real-world user feedback, while OneRec-Think incorporates explicit in-text reasoning \citep{deng2025onerec,zhou2025onerecv2,liu-etal-2026-onerec}.

Another line of work studies LLMs as rankers or instruction-tuned recommendation models. LLMRank evaluates LLMs as zero-shot rankers over candidate items \citep{hou2024large}, and TALLRec aligns LLMs with recommendation data through efficient instruction tuning \citep{bao2023tallrec}. Related studies analyze ChatGPT-style models for pointwise, pairwise, and listwise recommendation \citep{dai2023uncovering}, or use LLM-generated textual augmentations to improve personalized recommendation \citep{lyu2024llmrec}.

Despite their flexibility, many LLM-based recommenders rely on generative decoding, prompt-based ranking, or interaction-heavy inference, which can be expensive for large-scale retrieval. In contrast, our work revisits the efficient two-tower retrieval paradigm in the LLM era. Instead of using LLMs purely as generators or rankers, we use LLMs as semantic representation backbones and study how to improve two-tower retrieval through shared encoding, EOS pooling, cross-dataset transfer, and distillation from stronger teacher models.

\subsection{Neural Retrieval Models}

Neural retrieval models learn dense representations of queries and candidates in a shared embedding space, enabling efficient relevance estimation with an inner product, cosine similarity, or another lightweight matching function \citep{huang2013learning,reimers2019sentencebert,karpukhin2020dense}. In recommender systems, the corresponding two-tower architecture independently encodes user-side and item-side inputs. Its factorized structure permits item embeddings to be precomputed and indexed, making it well suited to large-scale candidate generation.

Neural retrieval builds on representation-learning approaches in collaborative filtering, including matrix factorization \citep{koren2009matrix}, Bayesian Personalized Ranking \citep{rendle2009bpr}, and Neural Collaborative Filtering \citep{he2017neural}. In information retrieval, DSSM introduced shared semantic representations learned from clickthrough data \citep{huang2013learning}, while Sentence-BERT and DPR established effective neural dual encoders for semantic similarity and open-domain retrieval \citep{reimers2019sentencebert,karpukhin2020dense}. For industrial recommendation, the YouTube deep recommendation model uses a neural candidate-generation stage to retrieve items from a large corpus before ranking \citep{covington2016deep}.

The efficiency of dual encoders comes at the cost of modeling fine-grained interactions between a user and an item. Cross-encoders jointly encode the pair and typically produce stronger ranking signals, but their inference cost prevents their direct use for first-stage retrieval over large item corpora~\citep{nogueira2019passage}. Late-interaction models provide an intermediate point on this efficiency--effectiveness spectrum by retaining token-level interactions at retrieval time \citep{khattab2020colbert}. Further work has improved dual-encoder training through sampling-bias correction, cross-batch negative sampling, and mixed negative sampling~\citep{yi2019sampling,wang2021crossbatch,yang2020mixed}; dual encoders can also benefit substantially from distillation from more expressive models~\citep{menon2022defense}.

Our work follows this neural retrieval paradigm while revisiting the two-tower architecture with modern LLM backbones. A shared LLM encodes the user history and item text in separate forward passes, producing representations that can be matched by a lightweight similarity function; item embeddings therefore remain precomputable and indexable. We further transfer cross-encoder ranking signals through distillation and refine only the user tower with latent reasoning, preserving the retrieval-time factorization. Compared with fully generative LLM recommenders, our approach retains the efficiency required for large-scale candidate generation while strengthening the semantic representations used for retrieval.

\end{document}